\documentclass[preprint2]{aastex}

\begin{document}

\title{New distances for a selected set of visual binaries with inconsistent dynamical masses}

\author{V. S. Tamazian\altaffilmark{1}, O. Yu. Malkov\altaffilmark{2}, J. A. Docobo\altaffilmark{1}, D. A. Chulkov\altaffilmark{2},  P. P. Campo\altaffilmark{1}}

\altaffiltext{1}{Observatorio Astron\'omico Ram\'on Mar\'{\i}a
Aller, Universidade de Santiago de Compostela, Avenida das Ciencias
s/n, 15782 Santiago de Compostela, Spain.vakhtang.tamazian@usc.es, joseangel.docobo@usc.es, pedropablo.campo@usc.es} \altaffiltext{2}{Institute of Astronomy of the Russian Academy of Sciences,
48 Pyatnitskaya Street, Moscow 119017, Russia. malkov@inasan.ru, chulkov@inasan.ru}

\begin{abstract}

We have selected a set of 17 visual binaries that demonstrate great
inconsistency between the systemic mass obtained through Kepler's
Third Law as compared to that calculated
 through standard mass-luminosity and mass-spectrum relationships.
A careful inspection of orbital data and parallaxes showed that the current orbits
of nine binaries (WDS 00155$-$1608, WDS 00174+0853, WDS 05017+2050, WDS 06410+0954,
WDS 16212$-$2536, WDS 17336$-$3706, WDS 19217$-$1557,
WDS 20312+1116, and WDS 21118+5959) do not need to be improved, instead we recommend
different parallax (distance) value for them. On the other hand,
we considered that eight orbits (WDS 02366+1227, WDS 02434-6643,
WDS 03244-1539, WDS 08507+1800, WDS 09128$-$6055, WDS 11532-1540, WDS 17375+2419,
and WDS 22408$-$0333) had to be improved.
Due to various reasons mentioned in this paper, their distances should most
likely be corrected unless better orbital solutions
and/or more precise parallaxes are reported. To obtain consistent
mass values, the use of the dynamical parallax is still recommended
 for 5 out of  the 8 improved orbits. For WDS 02434$-$6643,
WDS 09128$-$6055, and WDS 11532$-$1540, the improvement itself yields
reasonable mass sums while maintaining $\pi_{Hip}$
within a 1-2 $\sigma$ margin. New distance estimates for 16 stars (mainly based
on the obtained dynamical parallaxes)
and individual comments for all objects are presented and discussed.

\end{abstract}


\section{Introduction}

Visual binaries are a fundamental source of data on stellar masses as well as
a key observational interface for theoretical stellar evolution models.
The determination of accurate orbits in binary systems with well established parallaxes
represents a direct and reliable method for obtaining the dynamical mass of stars
through Kepler's Third Law, thus providing a useful constraint
on binary star formation and evolution mechanisms (Torres 2010, Mathieu 1994).

The direct application of Kepler's Third Law sometimes leads to an
anomalous dynamical mass for various reasons such as
poorly determined parallaxes and/or orbital elements, the existence of unknown companion(s), etc.
The differential photometry of pairs whose combined brightness is usually well known
allows us to estimate the luminosity and mass of individual components through
empirical mass--luminosity (M$-$L) and mass--spectrum relationships (Gray 2005, Schmidt-Kaler 1982).
Similarly, the accuracy of these relations is affected by various factors such as luminosity
effects, magnitude difference, variability, etc. (See Table 1).

In the paper of Malkov et al. (2012), dynamical masses were reported for a selected sample
of 652 visual binaries with essentially good quality orbital solutions,
along with estimated masses of their individual components. For a pair with
reasonably well-determined orbital elements, any large difference between these two sums
could possibly (but not only) indicate either an imprecise
or even erroneous parallax. On the other hand, such inconsistent data represent a
good starting point for a more detailed
overview and further improvement of the actual orbital elements.
Obviously, certain criteria should be applied in order to characterize the level
of inconsistency and to select the final set of objects as well.

In Section 2, we describe the requirements for the  binaries to be selected for further analysis.
We also explain the effect of the trigonometric parallax
on the observed inconsistency between dynamical and photometric masses.
Comments on particular systems along with brief presentations of their improved orbits and
masses are given in Section 3.
The obtained results are summarized in Section 4.

\section{The sample selection criteria}
\label{sec:select}

A careful inspection of dynamical masses for 652 binaries reported by Malkov et al. (2012)
with updated orbital data (as of September 2015) was carried out. When possible, the dynamical mass, $M_{d}$,
was first compared with the photometric mass, $M_{ph}$, estimated from
the observed photometry, the trigonometric parallax, and the M-L
relation and, secondly, with the spectral mass, $M_{sp}$ estimated from the spectral classification
and the mass-spectrum relation.

In most cases, only combined spectral types were available, and spectral type of the secondary was unknown. We
assigned individual spectral types following the method described by Edwards (1976)
applying magnitude
differences taken from the Fourth Catalog of Interferometric Measurements of Binary Stars
(Hartkopf et al. 2001a, hereinafter INT4).
Apart from this,  in our analysis we have made extensive use of dynamical mass, $M_{BR}$,
derived according to the
Baize \& Romani method (Baize \& Romani 1946; Heintz 1978) and the corresponding dynamical
parallax, $\pi_d$ (using updated M-L calibrations).

\begin{deluxetable}{lccc}
 \tablewidth{31pc}
\tablecaption{Influence of various parameters and stellar characteristics
on the value of dynamical, photometric, and spectral masses.}
\tablehead{\colhead{Parameter/reason}  & $M_{d}$ & $M_{ph}$ & $M_{sp}$ }
\startdata
Spectral (temperature) class &  no  & no     & yes  \\
Luminosity class             &  no  & yes    & yes  \\
Interstellar extinction      &  no  & yes    & no   \\
Trigonometric parallax       &  yes & yes    & no   \\
Semi-major axis, period      &  yes & no     & no   \\
Magnitude difference         &  no  & yes    & no   \\
Variability                  &  no  & yes    & (yes/no)\tablenotemark{a}  \\
Unresolved binarity          &  yes & yes    & no   \\
Third (undetected) component &  yes & no     & no   \\
Anomalous metallicity        &  no  & yes    & no   \\
Pre-MS stage                 &  no  & yes    & no   \\
\enddata
\tablenotetext{a}{\,Depends on variability type }

\end{deluxetable}

For photometric mass estimation, we used the M-L relation from Malkov (2007) for the upper-MS,
Henry \& McCarthy (1993) for the lower-MS, and Henry et al. (1999) for the lowest masses.
Subgiants and early-type (O$-$F6) giants were considered to be 1 mag brighter than
dwarfs (Halbwachs 1986).
Lastly, for the few remaining late-giant and supergiant stars, photometric masses
were estimated using
Tables II and VI of Straizys \& Kuriliene (1981). Pairs with unknown luminosity
class were considered to be MS-systems.
The luminosity class of the secondary component, when unknown, was considered to be
the same as for the primary.
If the magnitude of the secondary component was unknown, equal brightness of components
was assumed, and, consequently,
an upper limit for photometric mass was estimated. Spectral masses were estimated using
Table VI of Straizys \& Kuriliene (1981).
If the spectral type of the secondary was unknown, the listed spectral mass is the primary
mass (i.e., it represents
a minimum mass of the system). The main sequence is assumed if the luminosity class was
 unknown, as it was done in Malkov et al. (2012), which seems to be
a reasonable assumption for our relatively nearby orbital binaries.

In some cases, a relatively large discrepancy between the above mentioned masses was found.
The most obvious reasons for such discrepancies are imprecise parallaxes
and/or orbital elements as well as the presence of a third undetected body (or subcomponents)
that lead to the
$M_{d}$ overestimation (Tamazian et al. 2006) and a ``$M_{ph}$ excess'' for a given
luminosity (see, e.g., Kroupa et al., 1991; Malkov et al., 1997).
Among other reasons that explain the discrepancy between dynamical and photometric masses, one
should mention incorrect spectral classification,
interstellar extinction underestimation, and variability of the components. Individual
photometric masses also depend on stellar evolution
and chemical abundance variations (Bonfils et al. 2005). Finally, outdated cataloged
data and cross-identification errors can also distort mass values. The main reasons
that would lead to inconsistent dynamical mass values are listed in Table 1 and all
pairs with inconsistent masses require further study of their astrophysical
properties and dynamics, as well.

The accuracy of the dynamical mass calculated through Kepler's Third Law, $M_{d}$, mainly depends on
the quality of the orbital parameters and the parallax determination. The usual accuracy of an
MS star's photometric mass, $M_{ph}$, is about 15\%. Therefore, for a binary star, we assume
it to be $\approx$ 20\%. Out of the main sequence, the accuracy is 2-3 times lower and we assume
it to be $\approx$ 60\%. As for spectral mass, $M_{sp}$,
the accuracy is lower than 20\% and we assume 30\% accuracy for a binary.

For the systems where spectral classification is provided only for the primary component
and the magnitude difference is unknown, $M_{sp}$ should be considered as a lower limit.
In only one case (WDS 06410+0954), we found $M_{sp} > M_{ph}$. Other cases with obvious
disagreement (the difference between
various mass estimates exceeds double accuracy) were also considered. Additionally,
systems with unrealistic, high ($M_d>50M_\odot$)
and low ($M_d<0.2M_\odot$) dynamical masses were included.

The parallaxes are mainly taken from original (Perryman et al. 1997) and revised (van Leeuwen 2007)
reductions of the {\em Hipparcos} mission data.

Hipparcos parallaxes are subject to effects of stellar duplicity.
The companion was taken into account in calculating parallax,
when the pair is sufficiently wide to produce separate
measurements for the two components. At least in some cases
the presence of a companion increases the formal errors
(see van Leeuwen 2007 for discussion and examples).
It should be added that Hipparcos can provide unreliable
solutions for orbital periods close to one year, and
very long orbital periods can produce too insignificant
non-linear motions of the photocentre over the short (3-year)
measurement duration.

 The expected $M_{Ph}$ fits within the error bars
of $M_{Dyn}$ for 431 out of the 652 systems that comprise our sample (66\%). This number slightly increases
to 451 (69\%) if the original {\em Hipparcos} parallaxes are applied but it is mainly due to
their lower accuracy. If we consider both sets of the {\em Hipparcos} data ignoring errors, the
corresponding numbers become nearly identical (320 and 327 pairs when applying reprocessed and
original data, respectively) along with an obviously expected decrease in percentage (50\%).
Thus, when checking the consistency between dynamical and photometric masses in this large sample of visual
binaries, we have found no significant difference when applying either original or reprocessed {\em Hipparcos} data.

It is worth noting that $M_{dyn}$ tends to be larger than $M_{ph}$ for systems with a larger
error in the parallax, and $M_{dyn}>M_{ph}$ in 342 cases (52.4\%) for our entire sample.
If we apply the $\sigma_{\pi} < 20\%$ condition, the sample is reduced to 176 objects
and $M_{dyn}>M_{ph}$ for 84 of them (47.7\%). This means that we generally underestimate
stellar luminosity and, hence, $M_{ph}$ for systems with a larger error in the parallax (see discussion in Francis, 2014).

\begin{deluxetable}{lrcccrrcrrrcccc}
\tabletypesize{\tiny}
%\rotate
\tablecaption{Visual binaries with inconsistent dynamical mass and suggested parallaxes}
\tablewidth{0pt}
\tablehead{
\colhead{WDS} & \colhead{HIP} & \colhead{m$_1$} & \colhead{m$_2$} & \colhead{Sp1+Sp2} &
\colhead{$\pi_{Hip}$} & \colhead{$\sigma_{\pi_{Hip}}$} & \colhead{ $M_{d}$} &
\colhead{$\sigma_{M_d}$} & \colhead{$M_{ph}$} &
\colhead{$M_{sp}$} & \colhead{G} & \colhead{R} & \colhead{$\pi_{sugg.}$} & \colhead{$M_{BR}$}\\
\colhead{} & \colhead{} & \colhead{} & \colhead{} & \colhead{} &
\colhead{(mas)} & \colhead{(mas)} & \colhead{($M_{\odot}$)} &
\colhead{($M_{\odot}$)} & \colhead{($M_{\odot}$)} &
\colhead{($M_{\odot}$)} & \colhead{} & \colhead{} & \colhead{(mas)} & \colhead{($M_{\odot}$)}\\
\colhead{1} & \colhead{2} & \colhead{3} & \colhead{4} & \colhead{5} &
\colhead{6} & \colhead{7} & \colhead{ 8} &
\colhead{9} & \colhead{10} &
\colhead{11} & \colhead{12} & \colhead{13} & \colhead{14} & \colhead{15}
}
\startdata
00155$-$1608   & 1242&11.0  &11.4    &M3.5V+M5V   & 200.53   &  9.41  & 0.17   &  0.02  & 0.39   & 0.50  & 2     & dp     & 166.81\tablenotemark{a} & 0.29 \\
00174+0853     & 1392&7.78  & 8.38   &F7V+G0V     & 65.85    & 18.31  &  0.02  &  0.02  & 1.48   & 2.18  & 2     & d      & 15.31\tablenotemark{a}  & 1.48  \\
02366+1227**   & 12153&5.68  & 5.78   &F7V+F7V     & 28.79    & 0.43   & 19.1   &  2.57  & 2.95   & 2.26  & 1$_2$ &dp, ds   & 21.23\tablenotemark{b}  & 3.36 \\
02434$-$6643** & 12717&6.83  & 7.23   &F4V+F7V     & 18.40    & 0.43   & 14.6   &  4.04  & 2.67   & 2.38  & 3     &dp, ds   & 19.26\tablenotemark{c}  & 2.63 \\
03244$-$1539** & 15868&8.40  & 8.40   &G1V+G1V     & 20.27    & 0.73   &  0.62  &  0.11  & 2.1    & 2.0   & 3$_4$ &dp      & 28.89\tablenotemark{b}  & 1.64 \\
05017+2050     & 23396&8.45  & 9.32   &G0V+G5V     & 26.17    & 1.08   &  0.15  &  0.04  & 1.78   & 2.00  & 3     &dp, ds, d & 10.15\tablenotemark{b}  & 2.55 \\
06410+0954*    & 31978&4.66  & 5.90   &O7Ve+O9V    & 3.55     & 2.77   &  3.58  &  1.59  &17.3    &62.9   & 4     &ps      & 1.39\tablenotemark{a}   & 59.1\\
08507+1800**   & 43421&7.66  & 7.57   &G5V+G5V     & 2.77     & 1.03   &372.6   &461.5   & 7.8    & 1.90  & 3     & ps     & 34.37\tablenotemark{b}  & 1.80\\
09128$-$6055** & 45214&6.97  & 7.27   &B9V+B9V     & 5.77     & 0.48   & 34.1   & 12.5   & 5.4    & 5.1   & 3     & dp, ds  & 5.82\tablenotemark{c}   & 5.22\\
11532$-$1540** & 57955&8.61  & 9.29   &A9V+F1V     & 3.74     & 1.05   & 59.3   & 59.2   & 4.1   & 3.0   & 3     & d      & 5.77\tablenotemark{c}  & 3.07\\
16212$-$2536*  & 80112&3.30  & 4.10   &B1III+B1II  & 4.68     & 0.60   & 56.7   & 22.0   &15.6    &34.0   & 8     & d      & 5.76\tablenotemark{a}   & 30.3\\
17336$-$3706   & 85927&2.08  & 2.73   &B1.5IV+B2IV & 5.71     & 0.75   & 77.5   & 30.6   &22.6    &21.9   & 2     & d      &  8.93\tablenotemark{a}  & 18.5\\
17375+2419**   & 86254&5.90  & 7.30   &A1Vn+A8V    & 13.04    & 0.31   & 13.8   &  1.34  & 3.71   & 3.9   & 3     & dp, ds  & 18.40\tablenotemark{b}  & 3.06\\
19217$-$1557*  & 95176&4.58  & --     &F2p+F2      & 1.83     & 0.23   & 10.6   &  7.4   &32.1    &2.70   & 8     & ps     &  1.83\tablenotemark{a}  & 20.0\\
20312+1116     & 101233&7.90  & 8.00   &Am+Am       & 16.67    &10.67   &  0.20  &  0.38  & 2.46   & --    & 3     & d      & 7.43\tablenotemark{b}   & 3.93\\
21118+5959     & 104642&6.50  & 7.10   &B0II+B0II   & 0.76     & 0.42   &202.1   &352.2   &27.3    &50.2   & 3     & d      &  1.18\tablenotemark{a}  & 20.1\\
22408$-$0333** & 111465&6.52  & 8.63   &F6V+G9V     & 28.93    & 0.77   &  0.80  &  0.22  & 2.16   & 1.92  & 2$_3$ & dp     & 18.84\tablenotemark{b}  & 2.32\\
\enddata
%% Text for table notes should follow after the \enddata but before
%% the \end{deluxetable}. Make sure there is at least one \tablenotemark
%% in the table for each \tablenotetext.
\tablecomments{* spectroscopic binary;  ** newly calculated orbit; subscript in column 12 indicates
the orbit's previous grade}
\tablenotetext{a}{see comments in the text}
\tablenotetext{b}{$\pi_{d}$ }
\tablenotetext{c}{$\pi_{d}$ within 1-2 $\sigma$ of $\pi_{Hip}$ }
\end{deluxetable}

All of the selected systems are listed in Table 2.
The columns refer to: (1) the Washington Double Star Catalog (Mason et al. 2001, hereinafter WDS)
designation; (2) the {\em Hipparcos} identification; (3-4) the magnitudes and (5) the spectral types of
components; (6-7) the reprocessed {\em Hipparcos} parallax (van Leeuwen, 2007) and its
uncertainty; (8-9) the dynamical mass and its uncertainty; (10) the photometric mass; (11)
the spectral mass (all masses are given in units of solar mass); (12) the orbit quality grade
(G) according to the scheme described in the Sixth Catalog of Orbits of Visual Binary Stars
(Hartkopf et al. 2001b; hereinafter ORB6) where Grade 1 corresponds to a definitive orbit and
Grade 5 is assigned to an indeterminate orbit. For some improved orbits, the subscript indicates
their previous grade; (13) the reason or reasons, R, why the system is included in Table 2.
Among the reasons are the following: a large difference between
dynamical and photometric (dp), between dynamical and spectral (ds) and between photometric and
spectral (ps) masses, as well as a dynamical mass that is too high or too low (d).
The last two columns, 14 and 15,
list the suggested (mostly dynamical) parallax and the corresponding systemic mass, respectively.
Spectroscopic binaries are marked in Column 1 with an asterisk and pairs with newly
calculated orbits (Docobo et al. 2015) are marked with a double asterisk.

\section{Individual systems}
\label{sec:analysis}
In this section, we discuss individual binaries listed in Table~2.\\

%No. 1
{\bf WDS 00155$-$1608 (=HIP 1242).}
The application of the reprocessed {\em Hipparcos} parallax (200.53$\pm$9.41 mas)
to the orbit of Hershey \& Taff (1998) with a=0$\farcs$3037 and P=4.566 yr gives $M_{d}$=0.17 (hereinafter, all masses are given in units of solar mass), which is too low for a pair of M3.5+M5 dwarfs.
An astrometric study of these authors using HST FGS measurements
(last measurement dated 1997.964) allowed them to obtain a trigonometric
parallax of 166.6$\pm$0.8 mas and a reasonable mass sum of 0.29 (0.18+0.11).
Several high-quality astrometric measurements of this pair after 1997 were conducted
(Docobo et al., 2006; Tokovinin et al. 2015), and improved orbits were calculated (Perez et al. 2015; Tokovinin et al. 2015). However, the semiaxis (0$\farcs$3060) and the period (4.550 yr) of the latest solution given by Tokovinin et al. (2015) are very similar to those reported by Hershey \& Taff (1998), thus having small impact on the calculated total mass $M_{d}$=0.30 (instead of previous 0.29). Therefore,
the application of the parallax corresponding to a distance of 6.00 pc
(instead of 4.99 pc) seems to be more reasonable.
Notice that this value lies within a 2$\sigma$ margin of the original
{\em Hipparcos} parallax, 191.86$\pm$17.24 mas (Perryman et al. 1997).

%\end{document}
\vspace{1mm}
%No. 16
{\bf WDS 00174+0853 (=HIP 1392).}
The current orbit for this pair (P=35.7 yr, $a=0\farcs19$) has been reported by Hartkopf \& Mason (2010).
The use of the original {\em Hipparcos} parallax (15.31$\pm$1.35 mas; Perryman et al. 1997) instead
of that reprocessed by van Leeuwen's (2007) value of 65.85$\pm$18.31 mas readily
converts the Malkov et al. (2012) unrealistic dynamical mass of 0.02
to a reasonable $M_d=1.5\pm0.4$ for this pair of F7+G0 dwarfs.

%\end{document}
\vspace{1mm}
%No. 2
{\bf WDS 02366+1227 (=HIP 12153).}
The first, almost circular (e=0.037) orbit of Balega \& Balega (1988)
with a period of 3.87 yr and semi-major axis of a=0$\farcs$077  led
to $M_{d}$=3.8 (assuming $\pi_{Hip}$=28.8 mas)
which is large for a couple of F7 dwarfs. The solution of
Mason (1997) represents a highly elliptical orbit (e=0.88)
with a two-fold shorter period of 1.92 yr and a slightly larger
semi-major axis, a=0$\farcs$119, leading to an unrealistic $M_{d}$=19.1
while root mean squares (RMS) of (O$-$C) residuals both in $\rho$ and $\theta$ are lower
for this solution. We have calculated an improved orbit (Docobo et al. 2015; P=3.80 yr, a=0$\farcs$077, e=0.017)
which gives the lowest RMS value  and a reasonable mass of $M_{BR}$=3.4 ($\pi_d$=21.23 mas).

\vspace{1mm}
%No. 3
{\bf WDS 02434$-$6643 (=HIP 12717).}
Since a reliable magnitude difference, dm=0.4 mag, has been reported from speckle measurements (INT4 Catalog),
we did not apply dm=1.7 mag that appears in the ORB6 Catalog.
Using the combined magnitude, 6.26 mag, taken from SIMBAD,
we adopted 6.83 mag and 7.23 mag for the apparent brightness of the components.
From the improved orbit (Docobo et al. 2015), we obtained $\pi_d$=19.26 mas and $M_{BR}$=2.6 that are
compatible with F4+F7 dwarfs. Notice that $\pi_{Hip}$=18.40$\pm$0.43 mas.

\vspace{1mm}
%No. 4
{\bf WDS 03244$-$1539 (=HIP 15868).}
Previous orbits of Muller (1955) and Starikova (1978) have similar periods of about 25 yr, a=0$\farcs$15, and e=0.19
leading to $\pi_d\approx12.6$ mas ($\pi_{Hip}$=20.27$\pm$0.73 mas) and $M_d\approx0.6$ which
is inconsistent for a couple of early G dwarfs. A new, rather elliptical orbit by Docobo et al. (2015)
with a shorter period of 11.35 yr,
a larger semi-major axis (a=0$\farcs$17), and e=0.51, significantly improves the global RMS values.
Although its $\pi_d$=28.89 mas does not coincide well
with that of {\em Hipparcos}, we obtained a reasonable dynamical mass, $M_{BR}$, of about 1.6
and recommend this $\pi_{d}$ as a distance estimate.

\vspace{1mm}
%No. 6
{\bf WDS 05017+2050 (=HIP 23396).}
The current orbit (Hartkopf \& Mason 2010) is of good quality (Grade 3) and leads to
$\pi_d$=10.15 mas ($M_{BR}$=2.55), which is far from the $\pi_{Hip}$=26.17$\pm$1.08. We suggest the use of
$\pi_d$ as an estimate for its distance.

\vspace{1mm}
%No. 7
{\bf WDS 06410+0954 (=HIP 31978).}
These are the recently resolved subcomponents: Aa, Ab of CHR 168 (pre-main sequence
variable star, S Mon = 15 Mon). The dynamical mass, 3.58, given in Malkov et al. (2012) is unrealistic
for a couple of O7+09.5 dwarfs.
The reason for this is the application of an erroneous {\em Hipparcos} parallax of 3.58 mas (282 pc)
This value clearly differs from that used by Gies et al. (1997), 950 pc, as well as from the more probable
value of 720 pc suggested by
Cvetkovic et al. (2010). Both solutions lead to a similar dynamical mass (60)  that is compatible with the
spectral type. The orbit by Cvetkovic et al. (2010) has a significantly larger period
(P=74.3 yr) and a semi-major axis (a=0$\farcs$096) that is larger than the earlier
Gies' et al. (1997) solution (P=23.6 yr, a=0$\farcs$034). The RMS of residuals are smaller for
the Cvetkovic et al. (2010) and we recommend the use of a distance of 720 pc.

\vspace{1mm}
%No. 8
{\bf WDS 08507+1800 (=HIP 43421).}
The {\em Hipparcos} parallax, 2.77$\pm$1.03 mas, is clearly erroneous
because it leads to an unrealistic dynamical mass of 372
and an absolute magnitude of -0.3 mag for a G5V type star. The previous orbit (Hartkopf \& Mason, 2000)
with P=116.7 yr and a=0$\farcs$48 gave a $\pi_d$=13.82 mas
(72.4 pc) and a global mass of 3.0 which is large for a pair of G5 dwarfs.
We obtained a new orbit (Docobo et al. 2015)
with a similar period but larger semi-major axis (P=113.4 yr, a=0$\farcs$98)
providing $\pi_d$=34.37 mas (29.1 pc)
and reasonable values for both $M_{BR}$=1.8 and the absolute magnitude
of the main component (+5.5). Therefore, 29.1 pc is a realistic distance estimate for this star.

%\end{document}

\vspace{1mm}

%No. 9
{\bf WDS 09128$-$6055 (=HIP 45214).}
Previous orbits by Mason \& Hartkopf (2011)
and Heintz (1996) reported periods
of 78.5 yr (a=0$\farcs$343) and 71.3 yr (a=0$\farcs$265), respectively,
as well as dynamical masses of 34 and 19 which are too large for a pair of late B dwarfs.
A new orbit (Docobo et al. 2015) with a significantly larger period of P=400 yr
and a=0$\farcs$55 leads to $\pi_d$=5.82 mas and $M_{BR}$=5.2
which is quite reasonable for this pair.

\vspace{1mm}

%No. 20.
{\bf WDS 11532$-$1540 (=HIP 57955).}
The orbit by Hartkopf \& Mason (2010) with P=145.3 yr, a=0$\farcs$403, and e=0.287
yields $M_{BR}$=2.4 but its $\pi_d$=10.87 mas is then rather different from $\pi_{Hip}$=3.74$\pm$1.05 mas.
Docobo et al. (2015)  reported an almost two-fold larger period (P=236 yr),
rather elliptical (e=0.703) solution with a slightly smaller semi-major axis (a=0$\farcs$321)
that gives $M_{BR}$=3.1 which is more reasonable
for a pair of A9+F1 stars. Notice that its $\pi_d$=5.77 mas lies within a 2$\sigma$ margin of $\pi_{Hip}$.

\vspace{1mm}

%No. 19.
{\bf WDS 16212$-$2536 (=HIP 80112).}
This binary composed of two B giant stars belongs to the multiple system, $\sigma$~Sco.
Its combined visual-spectroscopic orbit
yields an orbital parallax of 5.76$\pm$ 0.68 mas (North et al. 2007). This value exceeds
that of {\em Hipparcos} (4.68$\pm$0.60 mas)
but it produces more reasonable masses of the components: 18.4$\pm$5.4 and 11.9 $\pm$ 3.1.
Tkachenko et al. (2014) calculated the masses to be 14.7$\pm$4.5 and 9.5$\pm$2.9 that
corresponds to an even greater parallax, 6.23 mas.

\vspace{1mm}

%No. 21.
{\bf WDS 17336$-$3706 (=HIP 85927).}
Tango et al. (2006) combined interferometric and spectroscopic data
for this pair and showed that it consists of B1.5IV and B2IV stars
located at $\pi_d$=8.93$\pm$0.4 mas
which is approximately 1.5 times larger than $\pi_{Hip}$=5.71$\pm$0.75 mas.

\vspace{1mm}
%No. 11
{\bf WDS 17375+2419 (=HIP 86254).}
There are two possible solutions for this system:
(i) a short-period, rather eccentric orbit ($P_1$=10.42 yr, 0$\farcs$127, e=0.75)
and (ii) a long-period, almost circular orbit ($P_2$=20.92, $0\farcs096$, e=0.03)
leading to $\pi_d$=18.34 mas and 7.63 mas, respectively. The $M_{BR}$ obtained
in the first solution (3.1) looks more realistic for a pair
of A1+A8 dwarfs than the second one (4.7),
and its $\pi_d$ is closer to $\pi_{Hip}$ (13.04$\pm$0.31 mas).

\vspace{1mm}

%No. 12
{\bf WDS 19217$-$1557 (=HIP 95176).}
This massive, single-lined spectroscopic binary
contains a bright star and an unseen companion.
In their medium spectral resolution interferometric observations,
Bonneau et al. (2011) found WDS 19217$-$1557 (= $\upsilon$ Sgr)
to be an interactive binary with a non-conservative evolution.
The B5-A0 supergiant status (rather than F2p, given in SIMBAD)
indicates $M_{sp}$ to be about 10 and makes the Malkov et al. (2012)
$M_{sp}$ estimations (1.35) incorrect. As the hotter, unseen secondary
component seems to be more massive (Bonneau et al. 2011),
the minimum total mass of the system should be at least $\approx$ 20
which is more consistent with the
Malkov et al. (2012), $M_{ph}$=32.1.
Yet, Bonneau et al. (2011) mention that the
separation of the stars may have been much smaller than estimated.
This hypothesis would imply that the distance to the system
is greatly underestimated. If so, $M_{d}$=10.6$\pm$7.4
(Malkov et al. 2012) should be increased.

\vspace{1mm}
%No. 18
{\bf WDS 20312+1116 (=HIP 101233).}
Tetzlaff et al. (2011) in their catalog of young
runaway {\em Hipparcos} stars give 1.5$\pm$0.1
for the mass of the primary which is consistent with
our $M_{ph}$=2.46 (Malkov et al. 2012). The unrealistic $M_{d}$
value (0.2) obtained by Malkov et al. (2012)
can be explained by either incorrect orbital
elements (the quality of the orbit is rather low) or an overestimated parallax:
van Leeuwen (2007) provided $\pi$=16.67$\pm$10.67 mas
used in Malkov et al. (2012) while the original {\em Hipparcos} value
is $\pi$=$-$3.06$\pm$9.53 mas (Perryman et al. 1997).
The current orbit of Hartkopf \& Mason (2014) leads to $\pi_d$=7.43 mas.

\vspace{1mm}

%No. 22
{\bf WDS 21118+5959 (=HIP 104642).}
According to the WDS discoverer designation, this system is McA 67Aa, Ab.
Its current orbital elements,  P=56.93 yr
and a=0$\farcs$066 reported by Zirm \& Rica (2012), along with a very poorly determined
$\pi_{Hip}=0.76\pm$0.42 mas led to an unrealistic mass of $M_{d}$=202.1. However,
applying the upper 1$\sigma$ limit for $\pi_{Hip}$ (1.18 mas), we obtained a
rather reasonable $M_{d}$=54 for a
couple of B0 giants. Notice that Stone (1978) computed a distance of
400 pc ($\pi$=2.5 mas) using the spectral
type and ``intrinsic color -- absolute magnitude -- spectral type''
calibrations, but it led to an unacceptably
small $M_{d}$=5.7 with a current orbit of Grade 3. A much more precise
parallax for this star is needed.
\vspace{1mm}

%No. 14
{\bf WDS 22408$-$0333 (=HIP 111965).}
A new orbital solution (Docobo et al. 2015; P=54 yr) improved the
previous orbit of Soderhjelm (1999) and agreed well with that of
Griffin \& Heintz  (1987) in radial velocity data.
We suggest using $\pi_d$=18.84 mas ($M_{BR}$=2.3) instead of $\pi_{Hip}$=28.93$\pm$0.77 mas.

%\end{document}

\section{Conclusions}
\label{sec:conclusions}

Of 652 visual binaries with essentially good quality orbits, a set of 17
pairs with largely
inconsistent (with standard mass-luminosity and mass-spectrum calibrations)
dynamical masses was selected.
The main results of this study can be summarized as follows:\\

%\begin{enumerate}

%\item

$\bullet$  On the basis of a careful overview of orbital and astrophysical data,
new distance estimates
(differing from those of {\em Hipparcos}) that restore the observed
dynamical mass consistency for 16 stars are suggested.\\

$\bullet$ We found no significant difference between original and reprocessed {\em Hipparcos} data
applications when detecting dynamical mass inconsistency.\\

$\bullet$ Recent orbital solutions for 8 binaries are briefly discussed. Three of them led us to reasonable
 dynamical masses while remaining within the 1-2 $\sigma$ margin of the {\em Hipparcos} parallax.\\

$\bullet$ Inconsistency of different mass estimates should be considered
 as an indicator for imprecise parallax and/or orbital elements of the binary system.\\

$\bullet$ $M_{ph}$ is generally underestimated, especially for systems with large parallax error.

\section{Acknowledgements}

%\begin{acknowledgements}
The authors thank the anonymous referee for useful comments.
This research has been supported by the Spanish Ministry
of Economy and Competitiveness under the Project AYA2011-26429.
The work was partially supported by the Program of Fundamental Researches of the Presidium of RAS (P-41),
by the Russian Foundation for Basic Research (projects 15-02-04053 and 16-07-01162), and
by the Program of the support of leading scientific schools of RF (3620.2014.2).
This research made use of the SIMBAD database
operated at the Centre de
Donn\'ees astronomiques de Strasbourg, the Washington Double Star Catalog maintained
at the U.S. Naval Observatory, and NASA's Astrophysics Data System Bibliographic Services.

%\end{acknowledgements}


\begin{thebibliography}{}

\bibitem{} Baize P. \& Romani L. 1946, Ann. d'Astrophys. 9, 13
\bibitem{} Balega Y. \& Balega I. 1988, Soviet Astron. Lettr. 14, 393
\bibitem{} Bonfils X., Delfosse X., Udry S. et al. 2005, A\&A 442, 635
\bibitem{} Bonneau D, Chesneau O, Mourard D. et al. 2011, A\&A 532, A148
\bibitem{} Cvetkovic Z., Vince I., \& Ninkovic S. 2010, NewA 15, 302
\bibitem{} Docobo, J. A., Tamazian, V. S., Malkov, O. Yu., Campo, P. P. \& Chulkov, D. I. 2015, IAU Commission 26 Circ. 185
\bibitem{} Docobo J. A., Tamazian V. S., Balega, Y. Y., \& Melikian N. D. 2006, AJ 132, 994
\bibitem{} Edwards T. W. 1976, AJ 81, 245
\bibitem{} Francis C. 2014, MNRAS, 444, L6
\bibitem{} Gies D. R., Mason B. D., Bagnuolo W. G. Jr. et al. 1997, ApJL 475, 49
\bibitem{} Gray D. F. 2005, in {\em The Observation and Analysis of Stellar Photospheres}, Cambridge Univ. Press: Cambridge, p. 506
\bibitem{} Griffin, R. F., \& Heintz, W. D.  1987, JRASC, 81, 3
\bibitem{} Halbwachs J.-L. 1986, A\&A 168, 161
\bibitem{} Hartkopf, W. I., Mason, B. D., Wycoff, G. L., \& McAlister, H. A. 2001a, Fourth Catalog of Interferometric Measurements of Binary Stars
\rm(http://www.usno.navy.mil/USNO/astrometry/optical-IR-prod/wds/int4)
\bibitem{} Hartkopf W. I., Mason B. D., \& Worley C. E. 2001b, Sixth Catalog of Orbits of Visual Binary Stars,
\rm{http://www.ad.usno.navy.mil/wds/orb6/orb6.html}
\bibitem{} Hartkopf W. I. \& Mason B. D. 2000, IAUDS 142, 1
\bibitem{} Hartkopf W. I. \& Mason B. D. 2010, IAUDS 170, 1
\bibitem{} Hartkopf W. I. \& Mason B. D. 2014, IAUDS 184, 1
\bibitem{} Heintz W. D. 1978, Double Stars, Dordrecht: Reidel, p. 62
\bibitem{} Heintz W. D. 1996, AJ 111, 412
\bibitem{} Henry T. J. \& McCarthy D. W. Jr. 1993, AJ 106, 773
\bibitem{} Henry T. J., Franz O. G., Wasserman L. H. et al. 1999, ApJ 512, 864
\bibitem{} Hershey J. L. \& Taff L. G. 1998, AJ 116, 1440
\bibitem{} Kroupa, P., Tout, Ch. A., \& Gilmore, G. 1991, 251, 293
\bibitem{} Malkov, O. Yu., Piskunov, A. E., \& Shpilkina, D. A. 1997, A\&A, 320, 79
\bibitem{} Malkov O. 2007, MNRAS 382, 1073
\bibitem{} Malkov O. Yu., Tamazian V. S., Docobo J. A. \& Chulkov D. A. 2012, A\&A 546, 69
\bibitem{} Mason B. D. 1997, AJ 114, 808
\bibitem{} Mason B. D. \& Hartkopf W. I. 2011, IAUDS 173, 1
\bibitem{} Mason B. D., Wycoff, G. L., Hartkopf, W. I., Douglass, G. G., \& Worley, C.E. 2001, The Washington Double Star Catalog, AJ, 122, 3466
\rm{http://www.usno.navy.mil/USNO/astrometry/optical-IR-prod/wds/WDS}
\bibitem{} Mathieu R. D. 1994, ARA\&A 32, 465
\bibitem{} Muller P. 1955, JO 38, 17
\bibitem{} North J. R., Davis J., Tuthill P. G. et al. 2007, MNRAS 380, 1276
\bibitem{} Perez, T., Mendez, R.A., \& Horch, E. P. 2015, IAUDS 186, 1
\bibitem{} Perryman M. A. C., et al. 1997, The \textit{Hipparcos} and Tycho Catalogues (ESA SP-1200; Noordwijk: ESA)
\bibitem{} Schmidt-Kaler T. 1982, in Landolt-B\"{o}rnstein New Series, Group 6, Vol. 2b, ed. K. Schaifers \& H.-H. Voigt ( Berlin: Springer), 18
\bibitem{} Soderhjelm S. 1999, A\&A, 341, 121
\bibitem{} Starikova G. A. 1978, Soviet Astron. Lettr. 4, 52
\bibitem{} Stone R. C. 1978, AJ 83, 393
\bibitem{} Straizys V. \& Kuriliene G. 1981, ApSS 80, 363
\bibitem{} Tamazian V. S., Docobo J. A., Melikian N. D. \& Karapetian A. A. 2006, PASP 118, 814
\bibitem{} Tango W. J., Davis J., Ireland M. J. et al. 2006, MNRAS 370, 884
\bibitem{} Tetzlaff N., Neuh\"{a}user R., \& Hohle M. M. 2011, MNRAS 410, 190
\bibitem{} Tkachenko A., Aerts C., Pavlovski K. et al. 2014, MNRAS 442, 616
\bibitem{} Tokovinin A., Mason B. D., Hartkopf W. I., Mendez, R. A., \& Horch, E. P. 2015, AJ 150, 50
\bibitem{} Torres G., Andersen J., \& Gim\'enez A. 2010, A\&A Rev. 18, 67
%\bibitem[]{} van Altena W. F., Lee J. T., \& Hoffleit E. D. 1995, The General Catalogue of Trigonometric Stellar Parallaxes, Fourth Edition, Yale University Observatory
\bibitem{} van Leeuwen F. 2007, A\&A 474, 653
\bibitem{} Zirm H. \& Rica F. 2012, IAU Commission 26 Circ. 176
\end{thebibliography}
\end{document}